\documentclass[aps,prb,twocolumn,amsmath,amssymb,superscript]{revtex4-1}
\usepackage{amsmath,amsfonts,amssymb,amsopn,mathtools}
\usepackage{bm,xcolor}
\usepackage[colorlinks=true,urlcolor=blue,linkcolor=blue,citecolor=blue]{hyperref}
\usepackage[capitalize]{cleveref}
\usepackage{graphicx,subfigure,multirow,diagbox}

\crefname{figure}{Figure}{Figures}

\newcommand{\figfile}[1]{#1.pdf}

\begin{document}
\title{Revealing excited states of rotational Bose--Einstein condensates}

\author{Jianyuan Yin$^{1,2}$}
\author{Zhen Huang$^{3}$}
\author{Yongyong Cai$^{4}$}
\author{Qiang Du$^{5}$}
\author{Lei Zhang$^{6,\star}$}
\affiliation{$^1$School of Mathematical Sciences, Laboratory of Mathematics and Applied Mathematics, Peking University, Beijing 100871, China.\\
$^2$Department of Mathematics, National University of Singapore, Singapore 119076, Singapore.\\
$^3$Department of Mathematics, University of California, Berkeley, California 94720, USA.\\
$^4$School of Mathematical Sciences, Beijing Normal University, Beijing 100875, China.\\
$^5$Department of Applied Physics and Applied Mathematics and Data Science Institute, Columbia University, New York, NY 10027, USA.\\
$^6$Beijing International Center for Mathematical Research, Center for Quantitative Biology, Center for Machine Learning Research, Peking University, Beijing 100871, China.\\
$^\star$Correspondence: {zhangl@math.pku.edu.cn}
}

\begin{abstract}
Rotational Bose--Einstein condensates can exhibit quantized vortices as topological excitations.
In this study, the ground and excited states of the rotational Bose--Einstein condensates are systematically studied by calculating the stationary points of the Gross--Pitaevskii energy functional.
Various excited states and their connections at different rotational frequencies are revealed in solution landscapes constructed with the constrained high-index saddle dynamics method.
Four excitation mechanisms are identified: vortex addition, rearrangement, merging, and splitting.
We demonstrate changes in the ground state with increasing rotational frequencies and decipher the evolution of the stability of ground states.
\end{abstract}

\maketitle

\section*{PUBLIC SUMMARY}
\begin{enumerate}
  \item Solution landscapes of rotational Bose--Einstein condensates are constructed to reveal ground and excited states.
  \item Four excitation mechanisms are identified: vortex addition, rearrangement, merging, and splitting.
  \item Evolution of the stability of different ground states is deciphered.
\end{enumerate}

\section*{INTRODUCTION}
Quantized vortices as topological defects play a crucial role in the study of superfluidity and superconductivity. \cite{matthews1999vortices, kawaguchi2012spinor}
These vortices have been the subject of extensive research in Bose--Einstein condensates (BECs) of degenerate quantum gases, both theoretically and experimentally. \cite{dalfovo1999theory,leggett2001bose,cornell2002nobel,bao2013mathematical}
A common scenario for generating these topological defects is when the system is stirred by rotating laser beams. \cite{caradoc1999coherent,madison2000vortex,abo2001observation,coddington2003observation,fetter2009rotating}
As the rotational frequency increases, the vortex structures experience successive complex topological changes. \cite{aftalion2001vortices,bao2005ground,cai2018vortex}

Theoretically, the problem of identifying stationary vortex states of rotational BECs can be effectively solved by finding the stationary points of the Gross--Pitaevskii (G--P) energy functional with an equality constraint corresponding to the mass conservation. \cite{bao2013mathematical}
The local nature of a stationary point is often described by its (Morse) index in Morse theory, \cite{milnor1963morse} i.e., the number of negative eigenvalues of the Hessian.
For example, a local minimizer has an index zero and can be computed using the imaginary time integration of the dynamic G--P equation.
A stationary state with a nonzero index would be a saddle point due to its unstable nature.
The global minimizer is often referred to as the \emph{ground state}.
Stationary points with higher values of energy, consisting of both local minimizers and saddle points, are called \emph{excited states}.

Contrary to many computational efforts for finding ground states, \cite{butts1999predicted,aftalion2005vortex,bao2004computing,bao2006efficient,edmonds2020vortex,takeuchi2021quantum,kasamatsu2003nonlinear, weiler2008spontaneous} there are no controllable search algorithms to systematically explore energy landscapes and compute excited states.
From well-chosen initial guesses, some excited states with certain symmetry can be obtained using Newton's methods, \cite{law2010stable,law2014dynamic} or deflated continuation algorithms. \cite{charalampidis2018computing}
The observations of vortex nucleation in BECs have been reported in many references, \cite{raman2001vortex,madison2001stationary,price2016vortex} which characterize transitions between ground states and metastable vortex states.
However, a global landscape of excited states of BECs remains largely unexplored.
Excitation mechanisms between different vortex states, which provide the dynamical pathway from a ground/excited state to another, are still a mystery.

In this article, we systematically examine excited states and excitation mechanisms of two-dimensional (2D) BECs trapped in an isotropic harmonic-oscillator potential within the framework of mean-field theory, assuming that no excitations are caused along the $z$ axis.
Specifically, we apply an efficient numerical method based on the constrained high-index saddle dynamics (CHiSD) to construct the solution landscape of the G--P energy.
The \emph{solution landscape} is a pathway map consisting of all stationary points and their connections, \cite{yin2020construction,yin2021searching} which provides an efficient approach to finding multiple stationary points  and their connections without tuning random initial guesses.
This methodology has been successfully applied to liquid crystals, \cite{han2021solution,shi2022nematic,wang2021modelling,yin2022solution} quasicrystals, \cite{yin2021transition} and diblock copolymers. \cite{xu2021solution}
Using the solution landscape approach, we reveal four excitation mechanisms of BECs, i.e., \emph{vortex addition, rearrangement, merging} and \emph{splitting}.
We further demonstrate how the ground state changes with increasing rotational frequencies and the evolution of the stability of the ground states.

\section*{MODELS AND METHODS}
\subsection*{Gross--Pitaevskii energy}
A stationary vortex state of 2D rotational BECs can be characterized as a stationary point of the G--P energy (a dimensionless form), \cite{bao2013mathematical}
\begin{equation}\label{eqn:becenergy}
E(\phi)=\int \left\{\frac{1}{2}|\nabla \phi|^{2}+V|\phi|^{2}+\frac{\beta}{2}|\phi |^{4}-\Omega \bar{\phi} L_z \phi\right\} \mathrm{d} \bm{x},
\end{equation}
on the unit sphere $\phi \in \mathcal{M} = \{\varphi\in L^2(\mathbb{R}^2, \mathbb{C}): E(\varphi)<\infty,\int |\varphi|^2 \mathrm{d}\bm x=1 \}$.
Here $\phi$ is the complex-valued wave function of BECs defined on $\mathbb{R}^2$ and $|\phi|^2$ represents the particle density.
$V(\bm x)$ is the trapping potential and $\beta$ characterizes the interaction rate.
$\Omega$ is the rotational frequency and $L_z = -\mathrm{i}(x\partial_y - y\partial_x)$ is the $z$-component of the angular momentum operator.
Equivalently, each stationary point solves a nonlinear eigenvalue problem,
\begin{equation}\label{eqn:eigen}
\mu\phi=\left(-\frac12 \nabla^2+V+ \beta|\phi|^2-\Omega L_z\right) \phi,
\end{equation}
where $\mu$ is the chemical potential calculated as
\begin{equation}\label{eqn:mu}
\mu=\int\left\{\frac{1}{2}|\nabla \phi|^{2}+V|\phi|^{2}+\beta|\phi |^{4}-\Omega \bar{\phi} L_z \phi\right\} \mathrm{d} \bm{x},
\end{equation}
or equivalently
\begin{equation}\label{eqn:mue}
\mu=E(\phi)+\int \frac{\beta}{2}|\phi |^{4} \mathrm{d} \bm{x}.
\end{equation}
In our numerical experiments, $V$ is taken as an isotropic harmonic oscillator $V(\bm x)=\frac12 |\bm x|^2$.
A strongly repulsive interaction regime is considered as $\beta=300$.
The physical units of parameters in this dimensionless model are well documented in the references.\cite{bao2013mathematical, aftalion2001vortices}

The (Riemannian) gradient of the G--P energy \ref{eqn:becenergy} writes as
\begin{equation}\label{eqn:riemanniangrad}
\begin{aligned}
&\operatorname{grad} E(\phi) =\mathcal{P}_{\phi}\nabla E(\phi) =2\mathcal{P}_{\phi}\left(\frac{\delta E}{\delta \bar{\phi}}\right)\\
=&~\mathcal{P}_{\phi}(-\nabla^2 \phi + 2V\phi+2\beta|\phi|^2\phi - 2 \Omega L_z \phi ).
\end{aligned}
\end{equation}
Here, $\mathcal{P}_{\phi}$ is the projection operator on the tangent space $T(\phi) = \{\psi: \langle \psi,\phi \rangle=0\}$, defined as
\begin{equation}\label{eqn:projection}
\mathcal{P}_{\phi}\psi = \psi - \langle\psi,\phi\rangle\phi.
\end{equation}
The inner product is defined as
\begin{equation}\label{eqn:innerproduct}
\langle\psi,\phi\rangle = \phi^\top\psi =\operatorname{Re}\left(\int \bar{\phi}\psi\mathrm{d} \bm{x}\right).
\end{equation}
$\hat{\phi}\in\mathcal{M}$ is a stationary state of BECs if the gradient vanishes, i.e., $\operatorname{grad} E(\hat{\phi})=0$, which is equivalent to the nonlinear eigenvalue problem in \ref{eqn:eigen}.

The (Riemannian) Hessian is, for $\nu\in T(\phi)$,
\begin{equation}\label{eqn:riemannianhess}
\begin{aligned}
&\operatorname{Hess} E(\phi)[\nu]=\mathcal{P}_{\phi}(\partial_\nu\operatorname{grad} E(\phi))\\
=&~\mathcal{P}_\phi (\nabla^2 E(\phi)\nu) - \langle \phi, \nabla E(\phi)\rangle \nu.
\end{aligned}
\end{equation}
The Hessian can be extended to the whole space as a self-adjoint operator $\operatorname{Hess} E(\phi)[\mathcal{P}_\phi\nu]$.
In numerical computations, we use central difference schemes to approximate spatial derivatives of $\phi$ in \ref{eqn:riemanniangrad} and \ref{eqn:riemannianhess}.

The index of a stationary state $\hat{\phi}\in\mathcal{M}$ is calculated as the number of negative eigenvalues of the Hessian $\operatorname{Hess} E(\hat{\phi})$.
Methods for partial eigenvalue problems, such as the locally optimal block preconditioned conjugate gradient method, \cite{knyazev2001toward} can be applied to calculate the index and corresponding eigenfunctions within a small computational cost.

Some invariance properties exist in the G--P energy.
From the radial symmetry of the isotropic trapping potential $V(\bm x)$, for a stationary state $\hat{\phi}\in\mathcal{M}$ and any $\vartheta\in\mathbb{R}$, a global phase translation $\mathrm{e}^{\mathrm{i}\vartheta}\hat{\phi}$ and a rotation around the origin $\hat{\phi}(x\cos\vartheta-y\sin\vartheta, x\sin\vartheta+y\cos\vartheta)$ are also stationary states with the same index and energy.
This property indicates that, in general, the Hessian at a stationary state has two zero eigenvalues.
As a special case, the states with one central vortex of a winding number $m$ or no vortex ($m=0$) can be expressed as $\mathrm{e}^{\mathrm{i}m\theta}\varphi_m(r)$ in polar coordinates $(r,\theta)$, so the Hessians at these stationary points have only one zero eigenvalue. \cite{bao2005ground,bao2013mathematical}
Different multiplicities of zero eigenvalues would bring some difficulties in numerical computations.
For some stationary states, Hessians have more zero eigenvalues, as explained in the \textcolor[rgb]{0.00,0.00,1.00}{supplemental information}, and consequently we obtain multiple numerical results of some states as shown in \textcolor[rgb]{0.00,0.00,1.00}{Figure~S1}.

\subsection*{CHiSD method}
Here we briefly introduce the CHiSD method and its numerical implementations to calculate the excited states, and the details can be found in the reference.\cite{yin2022constrained}
The CHiSD method can be regarded as a generalization of the imaginary time method to compute an index-$k$ saddle point ($k$-saddle).
For the sphere constraint, the CHiSD for a $k$-saddle ($k$-CHiSD) is,
\begin{equation}\label{eqn:chisd}
\left\{
\begin{aligned}
\dot{\phi}  =&-\left(\mathbf{I}-\sum\limits_{i=1}^{k} 2\nu_{i} \nu_{i}^{\top}\right) \operatorname{grad}E(\phi),\\
\dot{\nu}_i=&-\left(\mathbf{I}-\nu_i\nu_i^\top- \sum\limits_{j=1}^{i-1} 2\nu_j\nu_j^\top\right) \operatorname{Hess}E(\phi)[\nu_i]\\
&-\langle \nu_i, \operatorname{grad}E(\phi) \rangle \phi, \quad  i=1, \cdots, k,
\end{aligned}
\right.
\end{equation}
coupled with an initial condition satisfying
\begin{equation}\label{eqn:chisdini}
\phi\in\mathcal{M}, \; \langle\phi,\nu_i\rangle=0,\;\langle\nu_i,\nu_j\rangle=\delta_{ij}.
\end{equation}
We refer to some references \cite{yin2022constrained, zhang2022discretization} for CHiSD methods for general constraints and the numerical analysis, respectively.

With an initial condition satisfying \ref{eqn:chisdini}, the $k$-CHiSD of \ref{eqn:chisd} always satisfies the constraints in \ref{eqn:chisdini}.
The dynamics of $\phi$ in \ref{eqn:chisd} represents a transformed gradient flow on $\mathcal{M}$, which consists of the gradient ascent along tangent directions of $\nu_1,\cdots,\nu_k$, and the gradient descent along other orthogonal tangent directions.
Therefore, the dynamics attempts to maximize the energy only on a $k$-dimensional submanifold.
Meanwhile, the dynamics of $\nu_i$ in \ref{eqn:chisd} finds the normalized eigenvector corresponding to the $i$-th smallest eigenvalue of $\operatorname{Hess}E(\phi)$.

To ensure these constraints in numerical implementations, we utilize numerical tools of retractions and vector transport in manifold optimization. \cite{absil2008optimization}
A retraction $R_\phi$ moves $\phi\in\mathcal{M}$ along a tangent vector $\eta_{\phi}\in T(\phi)$ on the manifold to $R_\phi(\eta_{\phi})$.
When $\phi$ moves on $\mathcal{M}$ along a tangent vector $\eta_{\phi}\in T(\phi)$ characterized as $R_\phi(\eta_{\phi})$, the vector transport $\mathcal{T}_{\eta_{\phi}}(\nu_{\phi})$ gives how to numerically change a tangent vector $\nu_{\phi}\in T(\phi)$ to a tangent vector at $R_\phi(\eta_{\phi})$ accordingly, as a generalization of parallel translation.
There is a considerable amount of flexibility in how to choose the retraction and vector transport, while different choices may lead to different results.
In our numerical computations, we apply a retraction operator of
\begin{equation}\label{eqn:smretraction}
R_\phi(\eta_{\phi}) = \frac{\phi+\eta_{\phi}}{\|\phi+\eta_{\phi}\|},
\end{equation}
and a vector transport of
\begin{equation}\label{eqn:vectortransportsphere}
\mathcal{T}_{\eta_{\phi}}\nu_{\phi}=\nu_{\phi}- \dfrac{\langle\nu_{\phi}, \phi+\eta_{\phi}\rangle}{\|\phi+\eta_{\phi}\|^2}(\phi+\eta_{\phi}).
\end{equation}

With the retraction operator $R$ and the vector transport $\mathcal{T}$, the $k$-CHiSD in \ref{eqn:chisd} can be numerically implemented with the initial condition satisfying \ref{eqn:chisdini}.
We aim to calculate $\phi^{(n+1)}$ and $\nu_i^{(n+1)}$ at the $(n+1)$-th iteration step based on $\phi^{(n)}$ and $\nu_i^{(n)}$ in the previous step.
For the dynamics of $\phi$, we can implement an explicit scheme with a retraction as
\begin{equation}\label{eqn:xiteration}
\begin{aligned}
\phi^{(n+1)}&=
R_{\phi^{(n)}} (\alpha^{(n)} \eta^{(n)}), \\
\eta^{(n)}&=-\left(\mathbf{I}-\sum_{i=1}^{k} 2 \nu_{i}^{(n)} {\nu_{i}^{(n)}}^\top\right)
\operatorname{grad} E\left(\phi^{(n)}\right),
\end{aligned}
\end{equation}
to calculate $\phi^{(n+1)}$ with a step size $\alpha^{(n)}$.
Note that $\nu_1^{(n)},\cdots,\nu_k^{(n)}$ are orthonormal vectors in $T(\phi^{(n)})$, so $\eta^{(n)}$ lies in the tangent space $T(\phi^{(n)})$ and then $\phi$ remains in the manifold $\mathcal{M}$ due to retraction.
Then, the vector transport $\mathcal{T}_{\alpha^{(n)} \eta^{(n)}}$ at $\phi^{(n)}$ moves $\{\nu_i^{(n)}\}$ to $\{\tilde{\nu}_i^{(n)}\}$ in the tangent space $T(\phi^{(n+1)})$, which is characterized by the second term in $\nu_i$ dynamics.
The first term in $\nu_i$ dynamics aims to solve an eigenvalue problem
\begin{equation}\label{eqn:eigenproblem}
\begin{aligned}
&\min\limits_{\nu_i} \left\langle\operatorname{Hess}E(\phi^{(n+1)})[\mathcal{P}_{\phi^{(n+1)}} \nu_{i}], \nu_{i}\right\rangle,\\
&\mathrm{s.t.}\; \langle\phi^{(n+1)},\nu_i\rangle=0, \; \left\langle \nu_j,\nu_i\right\rangle=\delta_{ij},\; j=1, \cdots, i,
\end{aligned}
\end{equation}
using gradient flow.
Because the transported vector $\tilde{\nu}_i^{(n)}$ provides a good initial guess for this problem, we can apply one-step gradient descent to solve this in each iteration.
Generally speaking, vector transport may not maintain the orthonormality, so a Gram--Schmidt procedure is finally implemented to the obtained vectors.
The iteration is terminated if $\|\operatorname{grad}E(\phi^{(n)})\|$ is smaller than tolerance.

In numerical computations, the wave function $\phi$ is truncated into a bounded domain $D=[-8,8]^2$ with homogeneous Dirichlet boundary conditions on $\partial D$ because stationary states decay to zero exponentially in the far field due to the effect of the trapping potential $V(\bm x)$. \cite{bao2013mathematical}
The wave function $\phi$ is discretized using finite difference methods with $N=128$ nodes along each dimension.

\subsection*{Downward and upward search algorithms}
To identify excited states and excitation mechanisms, we combine the CHiSD method with the following downward search algorithm to construct the solution landscape.\cite{yin2020construction,yin2021searching}

\begin{figure}[htb]
\includegraphics[width=\linewidth]{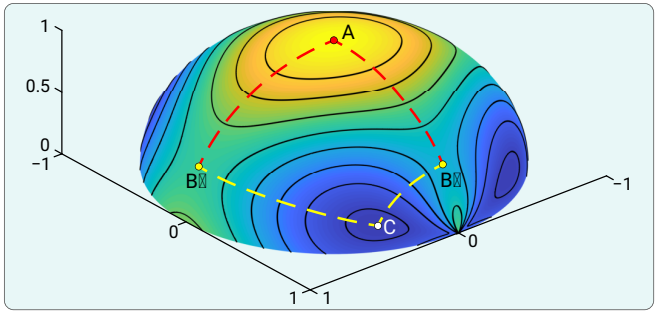}
\caption{\textbf{Illustration of the solution landscape on a unit sphere.}
Two 1-saddles $B_1$ and $B_2$ are connected to the index-2 stationary point $A$ (red dash lines).
The minimizer $C$ is connected to $B_1$ and $B_2$ (yellow dash lines).
The surface color from blue to yellow represents energy from low to high.
}
\label{fig:1}
\end{figure}

From an index-$k$ saddle point $\phi^*$ with orthonormal eigenvectors $\nu_1^*,\cdots,\nu_k^*\in T(\phi^*)$ corresponding to the negative eigenvalues of $\operatorname{Hess}E(\phi^*)$, we choose an unstable direction $\nu_i^*$ from them as the perturbation direction.
Then an $m$-CHiSD ($m<k$) is simulated from $R_{\phi^*}(\varepsilon \nu_i^*)$ with the initial directions $\{\mathcal{T}_{\varepsilon \nu_i^*}\nu_1,\cdots,\mathcal{T}_{\varepsilon \nu_i^*}\nu_m\}$, where $\nu_1, \cdots,\nu_m$ are chosen from the other unstable eigenvectors $\{\nu_1^*,\cdots,\nu_{i-1}^*, \nu_{i+1}^*,\cdots,\nu_k^*\}$.
Here a small constant $\varepsilon$, positive or negative, pushes the system away from the saddle point.
An orthogonal normalization procedure could be applied to the initial directions before simulation.
Different choices of $m$, $\nu_i^*$, and the sign of $\varepsilon$ may lead to different states.
By repeating this algorithm to newly-found states, we can finally reach different excited states and obtain connection relationships between states.
For the illustration example in \cref{fig:1}, from the index-2 stationary point $A$, two unstable directions lead to different 1-saddles $B_1$ and $B_2$ using 1-CHiSD.
Then the minimizer $C$ is obtained from either $B_1$ or $B_2$ by 0-CHiSD (i.e., gradient dynamics).

The downward search algorithm enables us to find multiple excited states from a high-index excited state.
On the other hand, we can also apply an upward search algorithm to find high-index excited states starting from a low-index excited state or a ground state, when the high-index excited state is unknown or multiple high-index excited states exist. \cite{yin2020construction}
Given an index-$k$ saddle point $\phi^\star$ with $l$ zero eigenvalues, we calculate the orthonormal eigenvectors $\nu_1^\star,\cdots,\nu_m^\star \in T(\phi^\star)$ corresponding to the smallest $m$ eigenvalues $\lambda_1^\star \leqslant \cdots\leqslant \lambda_m^\star$ of $\operatorname{Hess}E(\phi^\star)$, where $m>k+l$ so that $\lambda_m^\star>0$.
Then an $m$-CHiSD ($m>k$) is numerically simulated from $R_{\phi^\star}(\varepsilon \nu_m^\star)$, while the initial directions are reorthonormalization of $\mathcal{T}_{\varepsilon \nu_m^\star}\nu_1^\star,\cdots,\mathcal{T}_{\varepsilon \nu_m^\star}\nu_m^\star$.
Here $\varepsilon$ is also a small constant which could be positive or negative.
This algorithm can also be repeated to newly-found states.
In the illustration example in \cref{fig:1}, we may implement the upward search algorithm by using 1-CHiSD from a minimizer $C$ to obtain one of the index-1 saddle points $B_1$ and $B_2$.
Then the index-2 stationary point $A$ is achieved from this index-1 saddle point by applying 2-CHiSD.
A combination of the downward and upward search algorithms enables the entire search to navigate up and down on the solution landscape so that all excited states can be identified as long as they are connected somewhere.

\section*{RESULTS AND DISCUSSION}
\subsection*{Winding numbers}
To illustrate a vortex state $\phi$, the particle density $|\phi|^2$ of each state is plotted in $D$ within a common color bar.
Quantized vortices locate at points where $|\phi|^2=0$ and can be further classified according to their winding numbers.
The \emph{winding number} (topological charge) is an important feature of a quantized vortex, defined as how many times of $2\pi$ that the argument of $\phi$ changes around this vortex.
The ground state at $\Omega=0$, denoted as \textsf{O}, has no vortices.
To distinguish various vortex states, we denote ``\textsf{P}'' as a vortex of a winding number $+1$, and ``\textsf{N}'' as a $-1$ vortex.
The number of multiple vortices is attached as subscripts.
For example, \textsf{P}$_{\textsf2}$ represents two $+1$ vortices near the center.
For multiply vortices (high winding numbers), we denote ``\textsf{P}$^m$'' as a $+m$ vortex.

Because of the trapping potential $V$, the majority of particle density lies inside a circle $\{\bm x\in\mathbb{R}^2: \|\bm x\|=4\}$, while vortices of some states locate outside nearby this circle.
Therefore, we use ``\textsf{s}'' to separate vortices near the center and at the side.
For example, \textsf{P$_{\textsf2}$} has two $+1$ vortices near the center, while \textsf{sP$_{\textsf2}$} has two $+1$ side vortices near the circle.
\textsf{NP$_{\textsf4}$} has a $-1$ central vortex with four $+1$ vortices surrounded nearby, while four $+1$ vortices of \textsf{NsP$_{\textsf4}$} locate outside.

In the absence of rotation ($\Omega=0$), both \textsf{P} and \textsf{N} exist as 2-saddles.
With a positive frequency $\Omega$, vortices with positive winding numbers will be energetically favorable compared to those with negative winding numbers.

\begin{figure*}[hbtp]
\centering
\includegraphics[width=\linewidth]{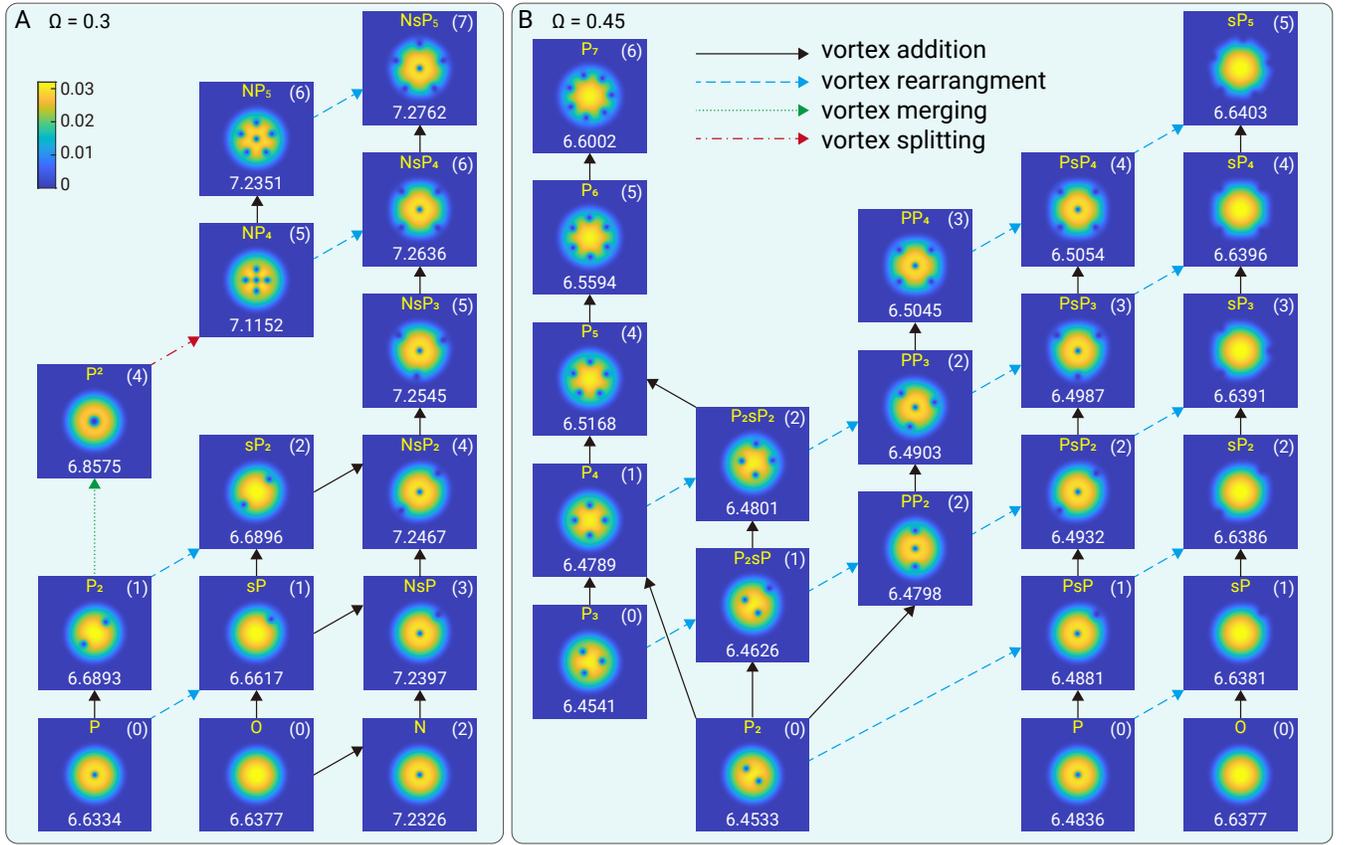}
\caption{\textbf{Excitation examples of four excitation mechanisms. }
(A) $\Omega=0.3$. (B) $\Omega=0.45$.
The four excitation mechanisms are vortex addition (black solid arrows), rearrangement (blue dash arrows), merging (green dotted arrow) and splitting (red dot dash arrow).
For each state here and after, its particle density $|\phi|^2$ is plotted in $D$ within a common color bar, while its name (top), energy (bottom), and index (top right parentheses) are labeled.
}
\label{fig:mec1}
\end{figure*}

\subsection*{Excitation mechanisms}
Without loss of generality, we demonstrate that excited states of BECs possess a variety of vortex structures at two frequencies, $\Omega=0.3$ and $0.45$, as illustrations.
Four excitation mechanisms are summarized by constructing solution landscapes.

\paragraph*{Vortex addition.}
The first and very common excitation mechanism is the vortex addition, i.e., adding new vortices from the far field.
As shown in black solid arrows of \cref{fig:mec1}A, this mechanism can be commonly observed in the $\Omega=0.3$ case, where \textsf{P} is the ground state and \textsf{O} is the first excited state.
Adding a side vortex of winding number $+1$ to \textsf{O} leads to an excited state \textsf{sP}, and further another excited state \textsf{sP}$_{\textsf2}$.
From the perspective of rare events, \textsf{sP} is the transition state (1-saddle) connecting \textsf{O} and \textsf{P}, and the unstable direction of \textsf{sP} corresponds to moving the vortex towards the center or outward.
This vortex addition mechanism can also be found in the excitation sequence \textsf{N} $\to$ \textsf{NsP} $\to$ \textsf{NsP}$_{\textsf2}$ $\to$ \textsf{NsP}$_{\textsf3}$ $\to$ \textsf{NsP}$_{\textsf4}$ $\to$ \textsf{NsP}$_{\textsf5}$ (see Video S1).
Along each excitation step, one $+1$ vortex is added from the far field to nearby the circle.

As shown in \cref{fig:mec1}B, more examples can be found at $\Omega=0.45$, each of which involves an additional $+1$ vortex from the far field.
In this case, \textsf{P}$_{\textsf2}$ and \textsf{P}$_{\textsf3}$ are the ground state and the first excited state, with similar energies.
From \textsf{P}$_{\textsf2}$, we can add side vortices as \textsf{P}$_{\textsf2}$ $\to$ \textsf{P}$_{\textsf2}$\textsf{sP} $\to$ \textsf{P}$_{\textsf2}$\textsf{sP}$_{\textsf2}$.
Once a side vortex is added, the index accordingly increases by one.
We can also add vortices around the central vortex to obtain \textsf{P}$_{\textsf2}$ $\to$ \textsf{PP}$_{\textsf2}$ $\to$ \textsf{PP}$_{\textsf3}$ $\to$ \textsf{PP}$_{\textsf4}$.
From \textsf{P}$_{\textsf3}$, we can successively add one middle vortex at a time to obtain excitations \textsf{P}$_{\textsf3}$ $\to$ \textsf{P}$_{\textsf4}$ $\to$ \textsf{P}$_{\textsf5}$ $\to$ \textsf{P}$_{\textsf6}$ $\to$ \textsf{P}$_{\textsf7}$, where vortices are arranged as regular polygons.
Compared to a side vortex, adding a middle one often increases the energy more significantly.
There are four local minimizers in this system, namely \textsf{O}, \textsf{P}, \textsf{P}$_{\textsf2}$, \textsf{P}$_{\textsf3}$, connected by three transition states \textsf{sP}, \textsf{PsP}, \textsf{P}$_{\textsf2}$\textsf{sP}, each of which has exactly one side vortex.
Along each minimum energy pathway, one $+1$ vortex is introduced from the far field to the middle.
Note that \textsf{P}$_{\textsf2}$\textsf{sP} has also been obtained as the transition state between \textsf{P}$_{\textsf2}$ and \textsf{P}$_{\textsf3}$ in reference \cite{schweigert1999flux}.

\paragraph*{Vortex rearrangement.}
Because of the confining trap, vortex positions significantly affect the energy.
At $\Omega=0.3$, the ground state \textsf{P} can be excited to the transition state \textsf{sP} by moving the central $+1$ vortex outward.
Similarly, \textsf{NP}$_{\textsf4}$ and \textsf{NP}$_{\textsf5}$ can also be excited to \textsf{NsP}$_{\textsf4}$ and \textsf{NsP}$_{\textsf5}$ respectively by rearranging the $+1$ vortices outward simultaneously, as shown in blue dash arrows of \cref{fig:mec1}A.

At $\Omega=0.45$, various vortex structures with the same number of vortices can be identified.
For \textsf{P}$_{\textsf2}$ with two vortices, moving one outward and the other to the center leads to a 1-saddle \textsf{PsP}, while moving both leads to \textsf{sP}$_{\textsf2}$.
For \textsf{P}$_{\textsf3}$ with three vortices, once a vortex is rearranged outside, the system is excited and the index increases by one.
Three vortices can also be aligned compactly as a 2-saddle \textsf{PP}$_{\textsf2}$ with a lower value of the energy than \textsf{PsP}$_{\textsf2}$.
Consequently, we obtain an excitation sequence \textsf{P}$_{\textsf3}$ $\to$ \textsf{P}$_{\textsf2}$\textsf{sP} $\to$ \textsf{PP}$_{\textsf2}$ $\to$ \textsf{PsP}$_{\textsf2}$ $\to$ \textsf{sP}$_{\textsf3}$ (see Video S2).
Similar excitations also occur for states with more vortices such as \textsf{P}$_{\textsf4}$.

\begin{figure}[htbp]
\centering
\includegraphics[width=\linewidth]{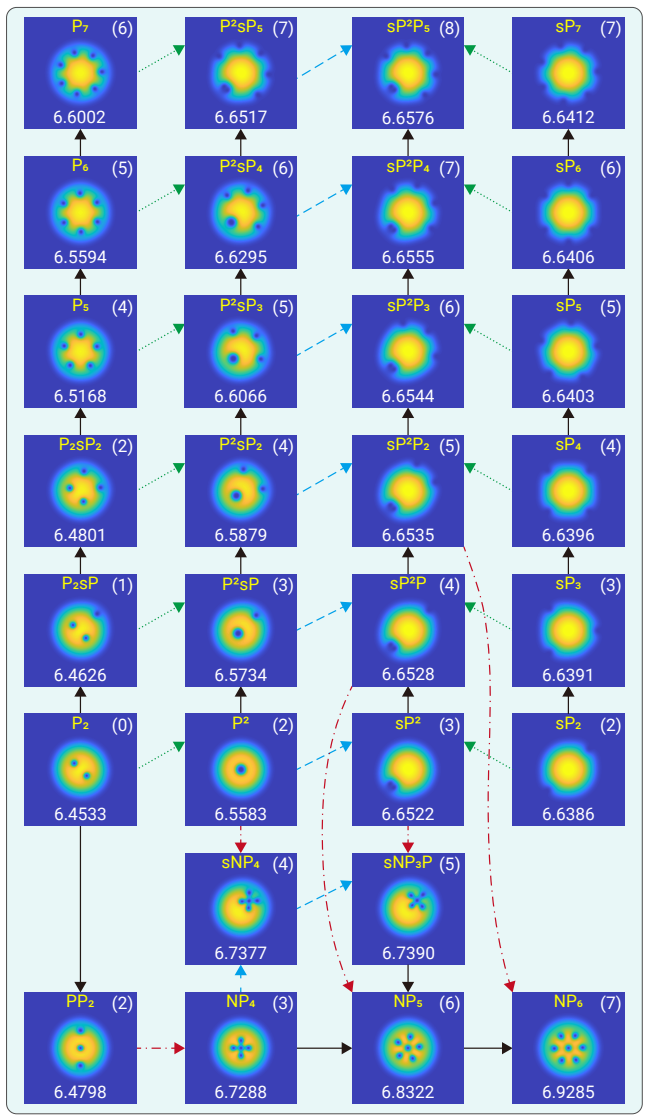}
\caption{\textbf{Excitation examples at $\Omega=0.45$.}
The legends are the same as \cref{fig:mec1}.
}
\label{fig:mec2}
\end{figure}

\paragraph*{Vortex merging.}
Stationary states presented above only involve vortices with winding numbers $\pm1$, while the BEC system also admits excited states with multiply vortices. \cite{aftalion2001vortices}
In fact, multiply vortices already exist in stationary states at $\Omega=0$, such as a central vortex state \textsf{P}$^{\textsf2}$.
Although one $+2$ vortex has the same winding number as two $+1$ vortices, this topological excitation often has a higher value of the energy and more unstable directions. \cite{kasamastu2009quantised}
At $\Omega=0.3$, merging two vortices of \textsf{P}$_{\textsf2}$ leads to \textsf{P}$^{\textsf2}$ with a significant energy increase, as shown in the green dotted arrow of \cref{fig:mec1}A.
\textsf{P}$^{\textsf2}$ is a 4-saddle with four unstable directions, two of which move the vortex outward, while the other two split it into two $+1$ vortices.

At $\Omega=0.45$, multiple states with $+2$ vortices can be obtained by merging.
Merging two vortices of \textsf{P}$_{\textsf2}$ also leads to \textsf{P}$^{\textsf2}$, which is a 2-saddle (see Video S3).
Along the unstable directions of \textsf{P}$^{\textsf2}$, the $+2$ vortex is split into two vortices, while moving the vortex outward leads to a 3-saddle \textsf{sP}$^{\textsf2}$, which can also be obtained by vortex merging from \textsf{sP}$_{\textsf2}$.
We can also add side vortices successively to \textsf{P}$^{\textsf2}$ with the index increasing by one for each side vortex, as shown in the excitation sequence \textsf{P}$^{\textsf2}$ $\to$ \textsf{P}$^{\textsf2}$\textsf{sP} $\to$ \textsf{P}$^{\textsf2}$\textsf{sP}$_{\textsf2}$ $\to$ \textsf{P}$^{\textsf2}$\textsf{sP}$_{\textsf3}$ $\to$ \textsf{P}$^{\textsf2}$\textsf{sP}$_{\textsf4}$ $\to$ \textsf{P}$^{\textsf2}$\textsf{sP}$_{\textsf5}$ with the $+2$ vortex going farther from the center.
These states can also be obtained by vortex merging from \textsf{P}$_{\textsf2}$\textsf{sP}, \textsf{P}$_{\textsf2}$\textsf{sP}$_{\textsf2}$, \textsf{P}$_{\textsf5}$, \textsf{P}$_{\textsf6}$, and \textsf{P}$_{\textsf7}$, respectively.
Similarly, \textsf{sP}$^{\textsf2}$\textsf{P}$_n$ ($n=1,\cdots,5$) can be obtained by successive vortex addition of \textsf{sP}$^{\textsf2}$, merging of \textsf{sP}$_{n+2}$, or rearranging \textsf{P}$^{\textsf2}$\textsf{sP}$_n$.
We enumerate these states in \cref{fig:mec2} with multiple excitation relations started from \textsf{P}$_{\textsf2}$ and \textsf{sP}$_{\textsf2}$.

\paragraph*{Vortex splitting.}
One vortex can also be split into multiple $\pm 1$ vortices as an excitation mechanism.
At $\Omega=0.3$, the $+2$ vortex of \textsf{P}$^{\textsf2}$ can be split into three $+1$ vortices and one $-1$ vortex.
With an addition of a fourth $+1$ vortex from the far field, the system is excited to \textsf{NP}$_{\textsf4}$, as shown in the red dot dash arrow of \cref{fig:mec1}A.

In the $\Omega=0.45$ case, similar excitation pathways can be identified.
from the 2-saddle \textsf{P}$^{\textsf2}$, the $+2$ vortex can also be split into three $+1$ vortices and one $-1$ vortex, leading to \textsf{sNP}$_{\textsf4}$ with an additional $+1$ vortex from the far field.
In a similar manner of vortex splitting, \textsf{sNP}$_{\textsf3}$\textsf{P}, \textsf{NP}$_{\textsf5}$ and \textsf{NP}$_{\textsf6}$ can also be obtained from \textsf{sP}$^{\textsf2}$, \textsf{sP}$^{\textsf2}$\textsf{P} and \textsf{sP}$^{\textsf2}$\textsf{P}$_{\textsf2}$, respectively.
Although vortex splitting does not change the total winding number, the additional vortices in these examples above do.
A $+1$ vortex can also be split into two $+1$ vortices and a $-1$ vortex with the total winding number unchanged.
From \textsf{PP}$_{\textsf2}$, the central vortex can split in this way and the system is excited to \textsf{NP}$_{\textsf4}$ (see Video S4), with no other vortices added.

\subsection*{Spectrum of stationary states}
Four excitation mechanisms are summarized above to generate excited states with different vortex structures.
Among these mechanisms, vortex addition will also change the total winding number.
Various excited state can be obtained using these mechanisms, and the CHiSD method can discover these excitation pathways.

Besides the excited states presented in \cref{fig:mec1} and \ref{fig:mec2}, we present the full spectrum of excited states found by the proposed method in \textcolor[rgb]{0.00,0.00,1.00}{Figure~S2} ($\Omega=0.3$) and \textcolor[rgb]{0.00,0.00,1.00}{S3} ($\Omega=0.45$).
A rich variety of vortex structures can be systematically obtained.
As the rotational frequency increases, the system can accommodate more positive vortices.
Note that theoretically an infinite number of excited states is expected in this system, so the upward search can be implemented continuously to obtain new excited states.
The presented results are not the entire family of stationary states.

Because the ground state and excited states are discussed in terms of energy, the stationary states are sorted in the ascending order of energy.
The chemical potential $\mu$ does not follow the same order, which is illustrated in \textcolor[rgb]{0.00,0.00,1.00}{Figure~S4}.
The stability based on the Morse index here is different from that of the Bogoliubov--de Gennes equation.
The latter addresses the real-time stability about the stationary states of the time-dependent G--P equation. \cite{dalfovo1999theory,fetter2009rotating,gao2020numerical}
Under the framework of the G--P energy and the CHiSD method, we can establish connections between different stationary states and calculate their transitions.

\begin{figure*}[htbp]
\includegraphics[width=\linewidth]{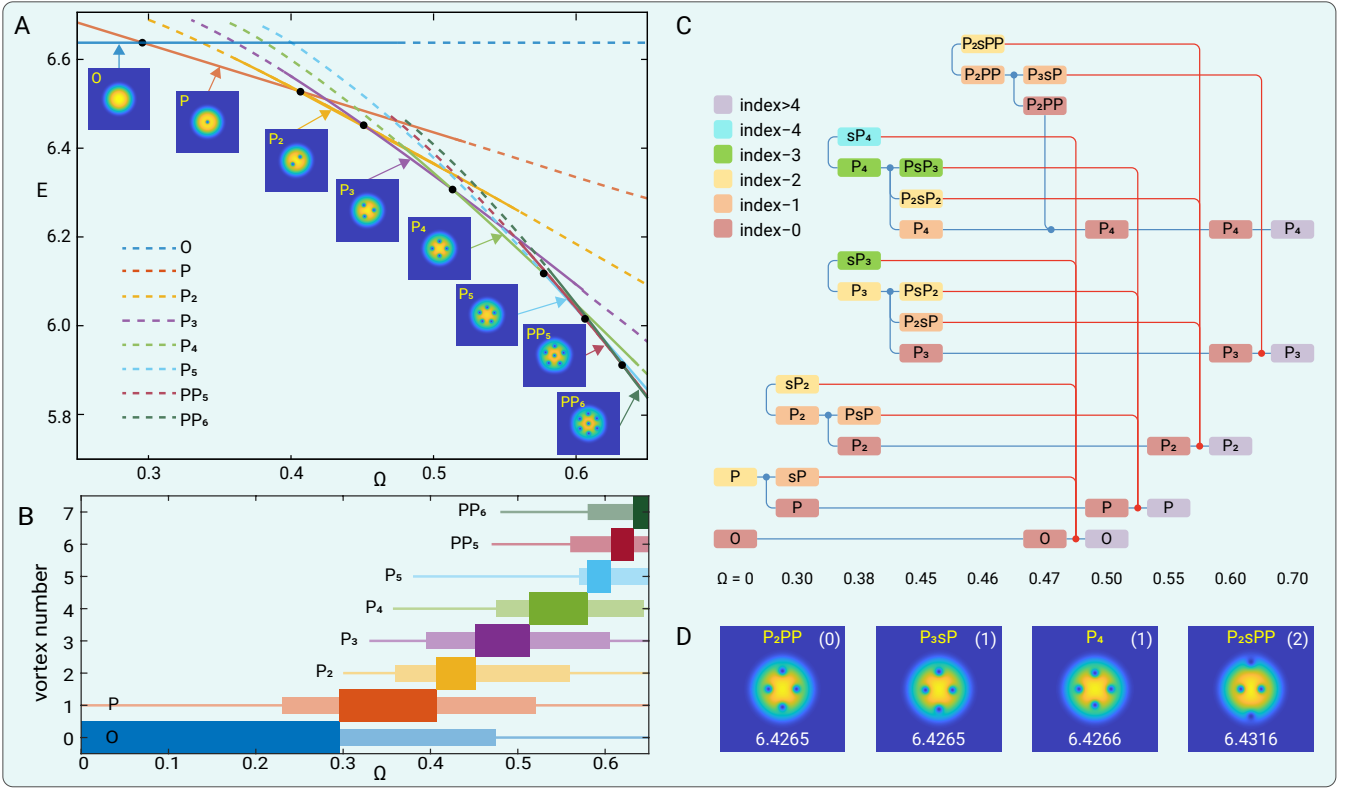}
\caption{
\textbf{Diagrams of the stability of different states. }
(A) An energy diagram for different $\Omega$.
As $\Omega$ increases, the ground state is respectively \textsf{O}, \textsf{P}, \textsf{P}$_{\textsf2}$, \textsf{P}$_{\textsf3}$, \textsf{P}$_{\textsf4}$, \textsf{P}$_{\textsf5}$, \textsf{PP}$_{\textsf5}$, \textsf{PP}$_{\textsf6}$.
Solid and dash lines represent local minimizers and saddle points respectively, while black dots indicate critical frequencies where the ground state changes.
(B) The stability of each state.
Solid rectangles indicates that the corresponding state is the ground state, while narrower light rectangles indicates that the corresponding state is a local minimizer and excited state. The solid line indicates where the state exists as a saddle point.
(C) Bifurcation diagram of some stationary states.
The color of each node represents its index.
\textsf{O} and \textsf{P} exist at $\Omega=0$, while \textsf{P}$_{\textsf2}$, \textsf{P}$_{\textsf3}$, \textsf{P}$_{\textsf4}$, and \textsf{P}$_{\textsf2}$\textsf{PP} emerge via saddle-node bifurcations.
Solid lines denote solution branches and T-junctions with dots denote pitchfork bifurcations.
Blue lines represent solution branches maintaining vortices inside, while red lines represent that some vortices move to the far field as $\Omega$ increases.
(D) Some states at $\Omega=0.47$ involved in the stabilization of \textsf{P}$_{\textsf4}$.
}
\label{fig:pdcom}
\end{figure*}

\subsection*{Phase diagram}
As the rotational frequency $\Omega$ exceeds a critical frequency, quantum phase transition takes place and the ground state has a topological change.
As a result, the vortex structure of the ground state experience successive topological changes with the increase of the frequency, as shown in the phase diagram in \cref{fig:pdcom}A and the stability illustration for each state in \cref{fig:pdcom}B.
For each ground state, the average position of vortices is located at the origin.
Similar vortex configurations and phase diagrams have also been reported in reference \cite{aftalion2001vortices}.

We start our discussions of the ground state evolution from \textsf{O}, the ground state at $\Omega=0$.
As $\Omega$ increases, \textsf{O} becomes a local minimizer (e.g. the first excited state at $\Omega=0.3$ and the 22nd excited state at $0.45$) and then a saddle point, while its energy remains almost unchanged.
The ground state at $\Omega=0.3$ is \textsf{P}, which exists as a 2-saddle at $\Omega=0$.
After a pitchfork bifurcation, \textsf{P} becomes a local minimizer with \textsf{sP} emerging as a 1-saddle.
Because \textsf{P} is a central vortex state, its Hessian has only one zero eigenvalue, while the Hessian at \textsf{sP} has two zero eigenvalues.
The different multiplicities of zero eigenvalues lead to index jumping of \textsf{P}.
As $\Omega$ increases, the energy of \textsf{P} continues decreasing as plotted in \cref{fig:pdcom}A, so that \textsf{P} replaces \textsf{O} as the ground state at the critical frequency. \cite{aftalion2001vortices}
Meanwhile, the 1-saddle \textsf{sP} moves its vortex far away and finally merges with \textsf{O} via a pitchfork bifurcation at $\Omega\approx0.47$, which makes \textsf{O} an unstable saddle point.

At $\Omega=0.45$, \textsf{P}$_{\textsf2}$ exists as the ground state.
In fact, \textsf{P}$_{\textsf2}$ and \textsf{sP}$_{\textsf2}$ do not exist in the $\Omega=0$ case, and emerge as saddle points via a saddle-node bifurcation at $\Omega<0.3$.
Then \textsf{P}$_{\textsf2}$ becomes a local minimizer via a pitchfork bifurcation, generating a 1-saddle \textsf{PsP}.
The two vortices of \textsf{P}$_{\textsf2}$ becomes closer, which accords with the results in Thomas--Fermi regime. \cite{aftalion2001vortices}
Because \textsf{P}$_{\textsf2}$ has more positive vortices than \textsf{P}, the energy of \textsf{P}$_{\textsf2}$ decreases faster as $\Omega$ increases, and \textsf{P}$_{\textsf2}$ replaces \textsf{P} as the ground state, while \textsf{P} exists as an excited state and local minimizer.
Meanwhile, the 1-saddle \textsf{PsP} moves its side vortex away and finally merges with \textsf{P} at $\Omega\approx0.53$, which then makes \textsf{P} an unstable saddle point.
As $\Omega$ increases, \textsf{P} goes through a sequence of ``saddle point $\rightarrow$ local minimizer $\rightarrow$ global minimizer $\rightarrow$ local minimizer $\rightarrow$ saddle point'', which exhibits as ``excited state $\rightarrow$ ground state $\rightarrow$ excited state''.
As shown in \cref{fig:pdcom}A, for a ground state at a higher frequency, which often accommodates more vortices, the energy decreases faster than the previous ground state with the increasing frequency.
Consequently, the ground states changes successively and the number of the vortices in the ground state also increases.
These results are qualitatively consistent with the theoretical predictions in the Thomas--Fermi regime. \cite{fetter2009rotating, aftalion2001vortices}

\subsection*{Bifurcation diagram}
To clearly illustrate how the excited states become stabilized, a bifurcation diagram is presented in \cref{fig:pdcom}C.
The change in the stability property of \textsf{P} is very generic for ground states at other $\Omega$, including \textsf{P}$_{\textsf2}$, \textsf{P}$_{\textsf3}$, and \textsf{P}$_{\textsf4}$.
After emergence, these states become more stable via some pitchfork bifurcations, and at the same time, some saddle points with central and side vortices are generated.
For each state after bifurcation, each side vortex near the circle brings an unstable direction, because moving it either outward or towards the center will decrease the energy, so its side vortex number coincides with its index.
Finally, as $\Omega$ increases, the side vortices move to the far field (illustrated with red lines in \cref{fig:pdcom}C, and the state merges with the corresponding local minimizer, leading to a high-index saddle point with central vortices.

The stabilization of \textsf{P}$_{\textsf4}$ is a little complicated because another state \textsf{P}$_{\textsf2}\textsf{PP}$ is involved.
\textsf{P}$_{\textsf2}$\textsf{PP} emerges as a 1-saddle first, and then is stabilized as a local minimizer via a pitchfork bifurcation, with a 1-saddle \textsf{P}$_{\textsf3}$\textsf{sP} emerging.
For a larger $\Omega$, \textsf{P}$_{\textsf4}$ is stabilized from a 1-saddle to a local minimizer via a pitchfork bifurcation involving this local minimizer \textsf{P}$_{\textsf2}$\textsf{PP}.
We present these states at $\Omega=0.47$ in \cref{fig:pdcom}D to illustrate this stabilization.

For a larger $\Omega$, the ground state would possess more vortices, and these vortices arrange themselves as multiple layers, as a result of the repulsive interactions between vortices and the attractive interactions between the vortices and the condensates.
For example, at $\Omega=0.62$ and $0.64$, respectively, \textsf{PP}$_{\textsf5}$ and \textsf{PP}$_{\textsf6}$ have a lower energy than \textsf{P}$_{\textsf6}$ and \textsf{P}$_{\textsf7}$.
For a fast-rotating condensate, the ground state can possess a complicated vortex lattice. \cite{aftalion2005vortex,aftalion2006vortex}
A regular vortex structure can appear in the ground state.

\section*{CONCLUSION}

This study provides a comprehensive and systematic examination of the vortex states and the excitation mechanisms in rotational Bose--Einstein condensates (BECs) trapped in an isotropic harmonic-oscillator potential.
Using a solution landscape approach by the CHiSD method combined with downward/upward search algorithms, the excited states of two-dimensional rotational BECs are revealed systematically.
Four distinct excitation mechanisms are identified from the results: vortex addition, rearrangement, merging, and splitting.
For each mechanism, we present a movie to illustrate the vortex behavior.
An excited state can be obtained along one or a few excitation pathways.
The changes in the ground state with increasing rotational frequencies are depicted using an energy diagram and explained through a bifurcation diagram.
As the rotational frequency increases, the ground state has more quantized vortices with a $+1$ winding number.
We also show that the ground state at a high rotational frequency actually emerges as an excited state at a low frequency first, and is then stabilized with the increase of the frequency.
For an overlarge frequency, this state becomes an excited state again.
The change in the stability property is generic for ground states at different frequencies.

The work can be naturally generalized to BECs trapped in an anisotropic harmonic-oscillator potential \cite{bao2005ground} to discover the corresponding excitation mechanisms.
For example, in the isotropic scenario, vortex addition along any direction from one minimizer \textsf{O} to another \textsf{P} corresponds to an identical energy barrier.
However, in an anisotropic scenario, vortex addition along different directions may result in different energy barriers, which could depend on the curvature of the anisotropic potential.

The methodology presented in this study offers an efficient numerical algorithm that constructs a complete solution landscape.
It serves as a powerful tool for solving a wide range of quantum systems, including self-attractive BECs, \cite{mihalache2006vortex} two-component BECs, \cite{hammond2022tunable} spinor BECs, \cite{kawaguchi2012spinor, chai2020magnetic} superconductors, \cite{schweigert1999flux,benfenati2020vortex} and fermionic wave functions. \cite{zhang2014optimal, zhang2016optimal, aoto2020calculating}

\section*{ACKNOWLEDGMENTS}
We thank Prof.~Weizhu Bao and Prof.~Biao Wu for the helpful discussions.
L.Z. is supported by the National Key Research and Development Program of China 2021YFF1200500 and the National Natural Science Foundation of China (No.12225102, T2321001, 12050002, and 12288101).
J.Y. is supported by the National Research Foundation, Singapore (Project No.~NRF-NRFF13-2021-0005).
Q.D. is supported by National Science Foundation (DMS-2012562 and DMS-1937254).
Y.C. is supported by the National Natural Science Foundation of China (No. 12171041).

\section*{AUTHOR CONTRIBUTIONS}
L.Z. and Y.C. designed research;
J.Y. and Z.H. performed research;
J.Y., Z.H., Y.C., Q.D. and L.Z. analyzed results;
J.Y., and L.Z. wrote the paper;
Y.C., Q.D. and L.Z. participated in manuscript revision.

\section*{DECLARATION OF INTERESTS}
The authors declare no competing interests.

\section*{SUPPLEMENTAL INFORMATION}
Supplemental information includes four figures and four movies.
\begin{enumerate}
  \item The excitation pathway sequence of vortex addition at $\Omega=0.3$: \textsf{N} $\to$ \textsf{NsP} $\to$ \textsf{NsP}$_{\textsf2}$ $\to$ \textsf{NsP}$_{\textsf3}$ $\to$ \textsf{NsP}$_{\textsf4}$ $\to$ \textsf{NsP}$_{\textsf5}$.
  \item The excitation pathway sequence of vortex rearrangement at $\Omega=0.45$: \textsf{P}$_{\textsf3}$ $\to$ \textsf{P}$_{\textsf2}$\textsf{sP} $\to$ \textsf{PP}$_{\textsf2}$ $\to$ \textsf{PsP}$_{\textsf2}$ $\to$ \textsf{sP}$_{\textsf3}$.
  \item The excitation pathway of vortex merging at $\Omega=0.45$: \textsf{P}$_{\textsf2}$ $\to$ \textsf{P}$^{\textsf2}$.
  \item The excitation pathway of vortex splitting at $\Omega=0.45$: \textsf{PP}$_{\textsf2}$ $\to$ \textsf{NP}$_{\textsf4}$.
\end{enumerate}

\section*{LEAD CONTACT WEBSITE}
\url{http://faculty.bicmr.pku.edu.cn/~zhanglei/}

\bibliographystyle{bibstyle}
\bibliography{bib}

\end{document}


\section*{Supplemental Information}

\section{Different numerical results of some states} \label{app:dif}
In the stationary states, vortices at different positions with respect to the trapping potential correspond to different values of the energy, and those closer to the center are often associated with lower values of the energy.
Interactions between vortices also contribute to the energy, so vortices tend to distribute uniformly for an equilibrium structure.
For example, the side vortex of \textsf{P}$_{\textsf2}$\textsf{sP} locates near the circle, equally far from the two middle vortices.

For two side vortices near the circle, interactions between them contribute much less to the energy, unless they get too close to each other.
For example, for \textsf{sP}$_n$ states, moving one vortex around the center only varies the energy up to the scale of $10^{-4}$, as long as no two vortices get too close.
Consequently, the Hessian at \textsf{sP}$_n$ has $(n+1)$ eigenvalues close to zero---one corresponds to a global phase factor $\mathrm{e}^{\mathrm{i}\vartheta}$, and the others correspond to moving $n$ side vortices around the center, while the Hessian at a general stationary state has two zero eigenvalues as stated in the main text.
This leads to multiple numerical results for each \textsf{sP}$_n$ state, which are treated as one excited state in practical computations.
Similar issues also happen to vortex states of \textsf{sP}$^{\textsf2}$\textsf{P}, \textsf{PsP} and \textsf{PsP}$_{\textsf2}$.
Some numerical results of stationary states at $\Omega=0.45$ are presented in \cref{fig:spn}.
Note that if any middle vortex deviates from the center, such as \textsf{P}$_{\textsf2}$\textsf{sP} and \textsf{P}$^{\textsf2}$\textsf{sP}, moving a side vortex will change the vortex distance.
Since the interactions between middle and side vortices vary for a different distance, the increase of energy cannot be neglected, so such an issue is absent.

This issue is resulted from the isotropic harmonic oscillator.
If we apply an anisotropic trap $V$, such a high degeneracy problem due to rotation invariance will no longer exist.
Moreover, one can also apply the Morse inequalities to estimate the number of stationary states of different indices in an anisotropic scenario.

\section{Spectrum of stationary states}
Here we present the full spectrum of excited states found by the proposed method in \cref{fig:bec030} ($\Omega=0.3$) and \cref{fig:bec045} ($\Omega=0.45$).

\section{Chemical potential of stationary states}\label{app:che}
The ground state and excited states are discussed in terms of the energy $E$, not the chemical potential $\mu$.
For linear cases $(\beta=0)$, $\mu$ coincides with $E$ as eigenvalues.
However, in this problem with strong nonlinearity, the chemical potentials of stationary states do not follow the same order of energies exactly.

We plot the energies and chemical potentials of some stationary states at $\Omega=0.3$ and $0.45$ in \cref{fig:emu}.
For example, at $\Omega=0.45$, the stationary state with the lowest chemical potential is the 1-saddle \textsf{P}$_{\textsf4}$, not the ground state.
As an interesting fact, adding $+1$ vortex aside, which can slightly increase the energy, often slightly decreases the chemical potential.
At $\Omega=0.3$, the chemical potential decreases along the excitation sequences \textsf{O} $\to$ \textsf{sP} $\to$ \textsf{sP}$_{\textsf2}$ as well as \textsf{N} $\to$ \textsf{NsP} $\to$ \textsf{NsP}$_{\textsf2}$ $\to$ \textsf{NsP}$_{\textsf3}$ $\to$ \textsf{NsP}$_{\textsf4}$ $\to$ \textsf{NsP}$_{\textsf5}$.
Similar issues happen at excitation sequences at $\Omega=0.45$ including \textsf{O} $\to$ \textsf{sP} $\to$ \textsf{sP}$_{\textsf2}$ $\to$ \textsf{sP}$_{\textsf3}$ $\to$ \textsf{sP}$_{\textsf4}$ $\to$ \textsf{sP}$_{\textsf5}$ $\to$ \textsf{sP}$_{\textsf6}$ $\to$ \textsf{sP}$_{\textsf7}$, \textsf{P} $\to$ \textsf{PsP} $\to$ \textsf{PsP}$_{\textsf2}$ $\to$ \textsf{PsP}$_{\textsf3}$ $\to$ \textsf{PsP}$_{\textsf4}$, and \textsf{sP}$^{\textsf2}$ $\to$ \textsf{sP}$^{\textsf2}$\textsf{P} $\to$ \textsf{sP}$^{\textsf2}$\textsf{P}$_{\textsf2}$ $\to$ \textsf{sP}$^{\textsf2}$\textsf{P}$_{\textsf3}$ $\to$ \textsf{sP}$^{\textsf2}$\textsf{P}$_{\textsf4}$ $\to$ \textsf{sP}$^{\textsf2}$\textsf{P}$_{\textsf5}$.
Note that the order of chemical potentials of \textsf{sP}$^{\textsf2}$, \textsf{sP}$^{\textsf2}$\textsf{P}, \textsf{sP}$^{\textsf2}$\textsf{P}$_{\textsf2}$, \textsf{sP}$^{\textsf2}$\textsf{P}$_{\textsf3}$, \textsf{sP}$^{\textsf2}$\textsf{P}$_{\textsf4}$, \textsf{sP}$^{\textsf2}$\textsf{P}$_{\textsf5}$ accords with that of \textsf{P}$_{\textsf2}$, \textsf{P}$_{\textsf2}$\textsf{sP}, \textsf{P}$_{\textsf2}$\textsf{sP}$_{\textsf2}$, \textsf{P}$_{\textsf5}$, \textsf{P}$_{\textsf6}$, \textsf{P}$_{\textsf7}$, because \textsf{P}$_{\textsf2}$ can only accommodate two side vortices at this frequency.

\begin{figure*}[p]
\includegraphics[width=\linewidth]{\figfile{figs1}}
\caption{\textbf{Different numerical results of some stationary states at $\Omega=0.45$.}
(A) \textsf{sP}$_{\textsf2}$. (B) \textsf{sP}$_{\textsf3}$. (C) \textsf{sP}$_{\textsf4}$. (D) \textsf{sP}$_{\textsf5}$. (E) \textsf{sP}$_{\textsf6}$. (F) \textsf{sP}$^{\textsf2}$\textsf{P}. (G) \textsf{PsP}$_{\textsf2}$. (H) \textsf{PsP}$_{\textsf3}$.
Each numerical solution $\phi$ satisfies $\|\operatorname{grad}E(\phi)\|<10^{-6}$.
For each group of states, their energies have a small difference of $\sim 10^{-4}$, so they are regarded as one stationary state in the main text.
}
\label{fig:spn}
\end{figure*}

\begin{figure*}[p]
\includegraphics[width=\linewidth]{\figfile{figs2}}
\caption{\textbf{Spectrum of stationary states at $\Omega=0.3$.}
}
\label{fig:bec030}
\end{figure*}

\begin{figure*}[p]
\includegraphics[width=\linewidth]{\figfile{figs3}}
\caption{\textbf{Spectrum of stationary states at $\Omega=0.45$.}
}
\label{fig:bec045}
\end{figure*}

\begin{figure*}[p]
\includegraphics[width=\linewidth]{\figfile{figs4}}
\caption{\textbf{Energies $E$ and chemical potentials $\mu$ of some stationary states.}
(A) $\Omega=0.3$. (B) $\Omega=0.45$.
Dashed arrows represent some states with successively increasing side $+1$ vortices.
For example, the dash arrow in (A) represents \textsf{N}, \textsf{NsP}, \textsf{NsP}$_{\textsf2}$, \textsf{NsP}$_{\textsf3}$, \textsf{NsP}$_{\textsf4}$, and  \textsf{NsP}$_{\textsf5}$.
}
\label{fig:emu}
\end{figure*}